\documentclass[12pt]{spieman}  
\usepackage{amsmath,amsfonts,amssymb}
\usepackage{algorithmic}
\usepackage{array}
\usepackage{graphicx}
\usepackage{makecell}
\usepackage{booktabs}
\usepackage{multirow}
\usepackage{setspace}
\usepackage{tocloft}
\usepackage[mathlines]{lineno}

\title{Deep Learning CT Image Restoration using System Blur and Noise Models}

\author[a]{Yijie Yuan}
\author[a,b]{Grace J. Gang}
\author[a, *]{J. Webster Stayman}
\affil[a]{Biomedical Engineering, Johns Hopkins University, Baltimore, MD, USA}
\affil[b]{Radiology, University of Pennsylvania, Philadelphia, PA, USA}

\cftpagenumbersoff{figure}
\cftpagenumbersoff{table} 
\cftpagenumbersoff{table} 
\begin{document} 
\maketitle

\begin{abstract}

\textbf{Purpose:} The restoration of images affected by blur and noise has been widely studied and has broad potential for applications including in medical imaging modalities like computed tomography (CT). Although the blur and noise in CT images can be attributed to a variety of system factors, these image properties can often be modeled and predicted accurately and used in classical restoration approaches for deconvolution and denoising. In classical approaches, simultaneous deconvolution and denoising can be challenging and often represent competing goals. Recently, deep learning approaches have demonstrated the potential to enhance image quality beyond classic limits; however, most deep learning models attempt a blind restoration problem and base their restoration on image inputs alone without direct knowledge of the image noise and blur properties. In this work, we present a method that leverages both degraded image inputs and a characterization of the system blur and noise to combine modeling and deep learning approaches. 

\textbf{Approach:} Different methods to integrate these auxiliary inputs are presented. Namely, an input-variant and a weight-variant approach wherein the auxiliary inputs are incorporated as a parameter vector before and after the convolutional block, respectively, allowing easy integration into any CNN architecture. 

\textbf{Results:} The proposed model shows superior performance compared to baseline models lacking auxiliary inputs. Evaluations are based on the average Peak Signal-to-Noise Ratio (PSNR), selected examples of good and poor performance for varying approaches, and an input space analysis to assess the effect of different noise and blur on performance.

\textbf{Conclusion:} Results demonstrate the efficacy of providing a deep learning model with auxiliary inputs, representing system blur and noise characteristics, to enhance the performance of the model in image restoration tasks.

\end{abstract}

\keywords{Computed Tomography, Images Restoration, Deep Learning}

{\noindent \footnotesize\textbf{*}J. Webster Stayman,  \linkable{web.stayman@jhu.edu}}

\begin{spacing}{2}   

\section{Introduction}
\label{sect:intro}  
Restoring images that have been contaminated by blur and noise has been widely studied and offers one potential pathway to improve diagnostic quality of medical images. Moreover, there is a long history of image quality modeling wherein the noise and spatial resolution properties of an image can be predicted\cite{BooneSeibert1994, fessler1996spatial, TwardSiewerdsen2008, gang2013modeling} or measured experimentally\cite{SolomonEtAl2012, Friedman2013, BUJILA201716}. With quantitative measures of image properties (e.g. the point spread function and the noise power spectrum), various restoration techniques including deconvolution and denoising can be applied. 


Numerous studies have been conducted on CT image deconvolution and denoising, with application to other medical imaging modalities as well as natural images. Traditional deconvolution methods encompass direct approaches like inverse filtering\cite{castleman1996}, Wiener filtering\cite{Hestenes1950TheEI, gonzalez2008digital}, and iterative methods such as Richardson-Lucy\cite{Lucy1974Iterative, Richardson1972Bayesian}, etc. Other related work on ''blind'' deconvolution\cite{giannakis2000, wang08, hosseini2019, aramarzi2013unified, xue2015novel, shah2017single, xiao2015stochastic, gregson2013stochastic, babacan2012bayesian, levin2009understanding, zuo2016learning, miao2022review, rani2016brief} seeks to restore an image without explicit information about the blur kernel and noise. One of the difficulties in many deconvolution methods is that deblurring generally magnifies noise. There has been a great deal of effort to develop sophisticated regularization schemes for model-based iterative approaches to help control noise magnification. Some common regularization techniques include image penalties based on total variation\cite{rodriguez2013total, chan1998total, osher2005iterative}, wavelet transforms \cite{shen2002image, neelamani2004forward}, non-local means\cite{buades2005nonlocal}, shrinkage fields\cite{schmidt2014shrinkage}, patch-based priors\cite{sun2014good, zoran2011learning}, Hyper-Laplacian priors\cite{krishnan2009fast}, and other fields of experts\cite{roth2005fields, schmidt2013discriminative, schmidt2011bayesian}, etc. 

More recently, deep-learning methods have been applied to solve the general image restoration problem. Many model architectures have been developed and used for image deconvolution and denoising tasks, including multi-layer perceptrons(MLPs)\cite{burger2012image}, convolutional neural networks(CNNs)\cite{sun2015learning, Tao_2018_CVPR, 7839189}, and transformers\cite{fan2022sunet, yao2022dense}. However, many of these approaches adopt a blind deconvolution approach without leveraging any prior knowledge of the blur and noise. In classic restoration approaches, blind deconvolution tends to be more difficult (more ill-conditioned) than restoring an image with known blur and noise distribution\cite{cai2009blind}. It is therefore reasonable to expect that including such system information would improve performance. One example of a processing scheme that combines deep learning with known image quality models is model-based deconvolution with a deep learning prior\cite{zhang2017learning, meinhardt2017learningproximal, arjomand2017deep, ulyanov2018deep, wang2019image, zukerman2020bp, gilton2019neumann}. There are potential drawbacks with this approach as most model-based approaches require iterative solution and relatively long processing times.


Since prior knowledge of blur and noise improves restoration performance in classical approaches, we hypothesize that the same is true for deep learning. Such approach has been applied to image deblurring where knowledge of the blur kernel is provided to the network. Ren et al. \cite{ren2018deep} initialized the first and third convolutional layers using a generalized low-rank approximation of the blur kernels. Guan et al.\cite{guan2022nonblind} locally shifted the low-dimensional layers (latent space) using auxiliary parameters that defined blur kernels. In this work, we develop a deep learning image restoration method that simultaneously perform deblurring and denoising while utilizing known information of the blur and noise distribution in the degraded image. Such information is encoded in auxiliary inputs provided to the network. 
Auxiliary inputs are commonly included in neural networks for conditional image generation \cite{mirza2014conditional, 8100115, 51403, ho2020denoising, karras2019style, zhang2023adding, wang2021ct, rombach2022highresolution, han2021dynamic} via concatenation or feature modification. In this work, we investigate methods for including such inputs for our particular application.    


We report below a deep learning image restoration scheme using CNNs, wherein the basic convolutional blocks were modified to encode auxiliary inputs to both the convolutional layer inputs and the intermediate feature map weights. These auxiliary inputs, which represent parameters defining the CT blur and noise kernels, can be generalized to any imaging system where the blur and noise can be accurately measured or predicted. This paper is distinct not only in incorporating auxiliary inputs but also in its more realistic modeling of blur and noise kernels compared to those using symmetrical Gaussians, and in conducting a thorough analysis of the input space. A preliminary investigation of this approach was previously reported\cite{yuan2023deep}. The following sections detail the development of these techniques and their evaluation in simulation experiments.

\section{Methods}
\label{sec:Method}
In this work, we seek to restore images that have been degraded according to the following forward model:
\begin{equation}
    \tilde{\mu} = h_1 **\ \mu + h_2 **\ n
\label{eq,forward}
\end{equation}
where $\mu$ represents a noiseless ground truth image that is degraded via convolution with blur kernel $h_1$ as well as additive correlated Gaussian noise (formed via convolution of a realization of white Gaussian noise, $n$, convolved with a kernel, $h_2$). We let $N$ denote the noise distribution associated with $h_2 **\ n$. We seek to obtain the restored $\mu$ given a noisy, blurred sample, $\tilde{\mu}$.
In a general supervised deep learning model, a mapping from the input space image $\tilde{\mu}$ to the output image $\mu$ is sought. 
Denoting $f(\theta)$ as a deep learning model with structure $f_\theta$ and parameters $\theta$, the optimization procedure for training the network can be written as:

\begin{equation}
f(\theta)=\underset{\theta}{\operatorname{argmin}}\ L\left(f_\theta(\tilde\mu_i, \alpha_{h_i}, \beta_{N_i}),\mu_i\right)
\label{eq,forward}
\end{equation}

where L is the loss function, and subscript $i$ denotes the index of different training samples. Auxiliary inputs $\alpha_{h_i}$ and $\beta_{N_i}$ are parameters that characterize $h_i$ and $N_i$. Note that this allows one to train across a range of different image properties by providing a range of $\alpha_{h_i}$ and $\beta_{N_i}$ samples. The particular forms of $\alpha_{h_i}$ and $\beta_{N_i}$ should be chose to represent the particular variability in blur and noise one would want to restore. For example, $\alpha_{h_i}$ could be the entire kernel $h_i$ or a lower dimensional representation of a large class of blur functions - e.g., the full-width half-maximum of a Gaussian blur function. 

\subsection{Network architecture with auxiliary inputs}

\begin{figure*}[t!]
\centering
\includegraphics[width=6.5in]{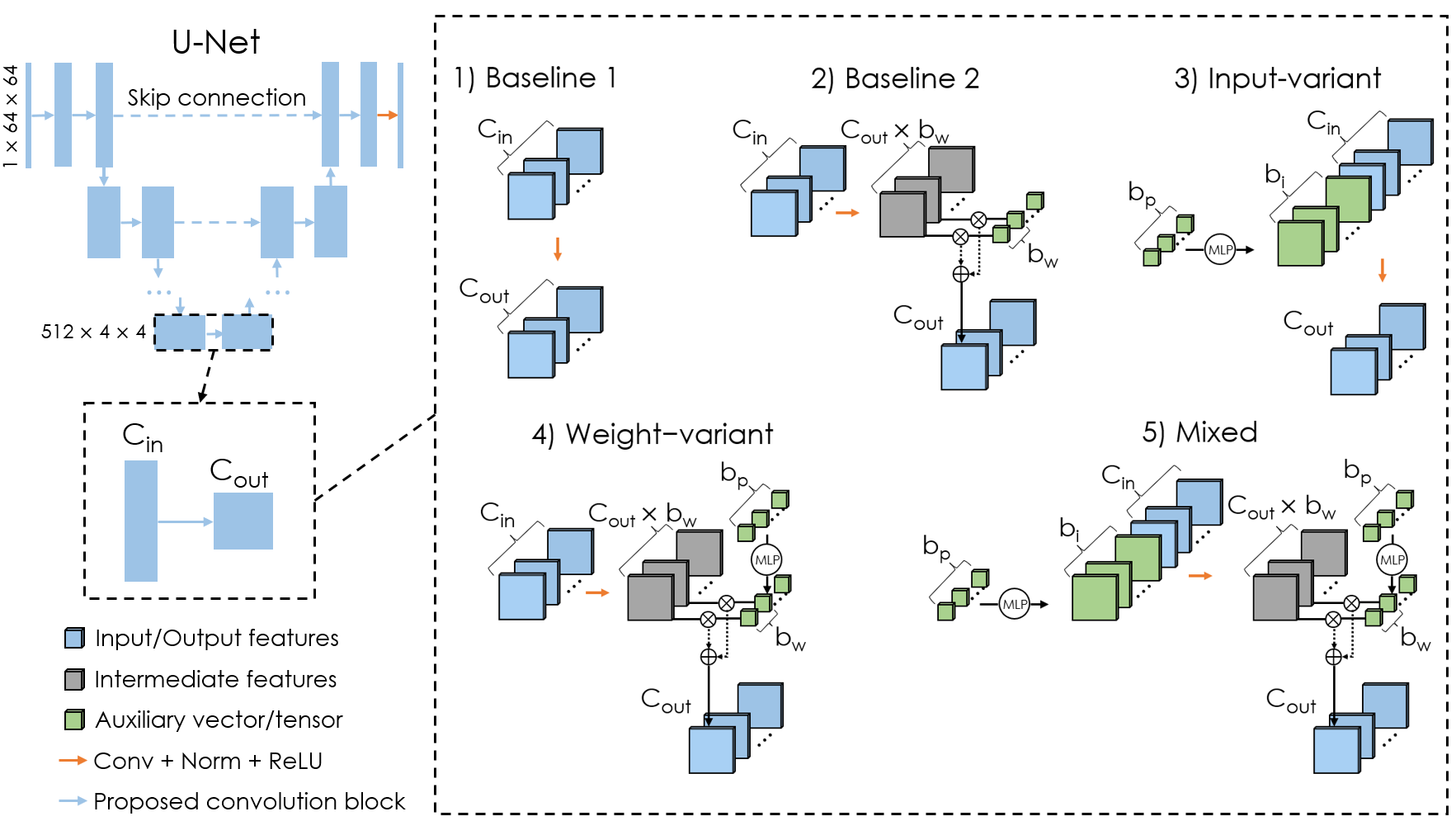}
\caption{Modified Model Structure. Our suggested changes focus on the convolutional block denoted by the black dashed rectangle, introducing three variations and two baselines (indicated by the blue arrow) to replace the conventional convolutional block (indicated by the orange arrows), with the exception of the final output layer where the standard block is retained. The input-variant and weight-variant models incorporate the auxiliary input prior to and following the convolutional process, respectively, while the mixed model utilizes it in both positions.}
\label{fig:architecture}
\end{figure*}

In this study, we form a new model architectures based on a variant of the U-Net\cite{ronneberger2015unet}\cite{isensee2018nnu}. The U-Net is widely used in medical image segmentation\cite{yin2022unet} and cross-modality translation\cite{Yang2020} tasks that ensures the same size of input and output. 

To integrate auxiliary image properties into the model, we introduce three approaches: input-based, weight-based, and a mixed method. The primary modifications involve the convolutional blocks, which are the building blocks of U-Net, as illustrated in Figure~\ref{fig:architecture}. Specifically, we convert standard convolutional blocks (Conv-Norm-ReLU operation) with $C_{in}$ input channels and $C_{out}$ output channels, to modified blocks that incorporate auxiliary vectors ($\alpha_{h_1},\beta_{h_2}$). Specifically, we modify {\em all} blocks except for the final output layer with these proposed modified blocks. We consider different block variations where the auxiliary vectors are either concatenated with the input (input-based) or serve as a weighting factor post-convolution (weight-based). These integrations enhance the model's ability to capture spatial and contextual information, thereby improving its performance. A detailed description of each approach follows:

\subsubsection{Input-variant}
In the input-based method, auxiliary vectors are concatenated with the convolutional block input, enabling the model to directly integrate additional information. As depicted in Figure 1, an auxiliary vector of length $b_p$ consists of scalars representing image properties (i.e., elements of $\alpha_{h_1}$ and $\beta_{h_2}$). This vector is fed into a multi-layer perceptron (MLP) to generate a new vector of length $b_i$. This step allows the model to learn relationships among scalars and utilize the combined information more efficiently. Each scalar in the vector is then transformed into a constant channel with equal pixel values as the scalar and the same size as the input feature channel. These $b_i$ constant channels form a constant block, which is concatenated with the original input before convolution. The number of constant channels, $b_i$, is an adjustable hyperparameter.
  
\subsubsection{Weight-variant}
In the weight-variant approach, the auxiliary vectors are used as weightings after the Conv-Norm-ReLU operation, allowing the model to adjust the importance of different features based on the auxiliary information. In this modification to the baseline U-Net architecture, the number of input and output channels remains unchanged. However, the number of output channels for the convolutional block is initially increased by a factor of $b_w$. These expanded channels are then divided into  $C_{out}$ groups, each containing $b_w$ channels. Then, we use a weighting vector, generated by a MLP from the auxiliary vector, to calculate the weighted average for each group of channels. This process produces one channel per group, and with $C_{out}$ groups, it yields $C_{out}$ channels in the block's final output. 

$$
C'_j=\sum_{i=1}^{b_w}w_{ij}\times C_{ij}
$$
where j is the index of the output channel which starts from 1 to $C_{out}$, $C'_j$ is the $j^{th}$ shifted channel, $w_{ij}$ is the $(b_w \times j + i)^{th}$ element in the encoded kernel vector, and $C_{ij}$ the $(b_w \times j + i)^{th}$ channel in the output of the convolution layer, respectively. The number of elements in the weighting vector $b_w$ is a tunable parameter as well.
  
\subsubsection{Mixed model}
The input-variant and weight-variant method consists of integrating auxiliary information both prior and after the convolution process. However, we can also consider a mixed approach which leverages the strengths of both methods. In this mixed model, we add the auxiliary input both prior to and after the convolution, following the principles of the input- and weight-variant approach. This enables the auxiliary information to play a more integral role in the network. As a result, this could potentially improve the network's performance.

\subsubsection{Baselines for Comparison}
In addition to the aforementioned modified block with auxiliary inputs, we describe two ``baseline'' blocks for comparison. 
For the input-based method, we trained a U-Net without incorporating auxiliary information or additional channels (Baseline 1). However, the weight-based and mixed models introduce approximately $b_w$ times more parameters, permitting additional computational complexity. For a fair comparison, we trained a second baseline (Baseline 2) in which the MLP-generated weighting vector was replaced by a trainable vector but without auxiliary inputs. This ensures that the only distinction between the proposed model and our baseline is whether the auxiliary vector is trainable or derived from auxiliary parameters, rather than computational complexity differences with original, baseline U-Net.


\begin{table*}[!t]
\centering
\renewcommand\arraystretch{1.5}
\caption{Three restoration scenarios of varying complexity}
\resizebox{1\textwidth}{!}{%
\begin{tabular}{|c|c|c|c|c|}
\hline
\textbf{Scenario} & \textbf{Forward Model} & \textbf{Parameter \& Range} & $\alpha_{h_1}$ & $\beta_{h_2}$ \\
\hline
Variable Blur & $\tilde{\mu} = h_1(\sigma_{h_1}) * \mu + n(\sigma_n)$ & $\sigma_{h_1} \sim U(0, 10), \sigma_n = 0.1$ & $\sigma_{h_1}$ & \\ 
\hline
Variable Noise & $\tilde{\mu} = \mu + h_2(\sigma_{h_2}) * n(\sigma_n)$ & $\sigma_{h_2} \sim U(0, 10), \sigma_n \sim U(0, 0.5)$ && $\sigma_{h_2},\sigma_n$\\ 
\hline
Variable Blur and Noise & \makecell{%
$\tilde{\mu} = h_1(M_1,E_1,O_1) * \mu$ \\ 
$+ h_2(M_2,E_2,O_2) * n(\sigma_n)$} & \makecell{%
$M_{1,2} \sim U(0, 10), E_{1,2} \sim U(0, 3),$ \\
$O_{1,2} \sim U(0, 180^\circ), \sigma_n \sim U(0, 0.2)$} & $M_1,E_1,O_1$ & $M_2,E_2,O_2,\sigma_n$\\
\hline
\end{tabular}%
}
\label{tab:scenarios}
\end{table*}

\subsection{Three Restoration Scenarios}
\label{Auxiliary input}
In this work, we consider three different restoration scenarios of varying complexity. These cases including the specific forward model (all are special cases of equation~(\ref{eq,forward})), parameters, ranges, and auxiliary inputs are summarized in Table~\ref{tab:scenarios} and explained in detail below.

\subsubsection{Variable Blur}
In this case, we consider the case of a simple parametric blur applied to an image with uncorrelated Gaussian noise. The blur function is a 2D symmetric Gaussian function with a width defined by $\sigma_{h_1}$, which is varied from 0 to 10. The magnitude of noise is fixed with a standard deviation equal to $\sigma_n = 0.1$.

\subsubsection{Variable Noise}
For this scenario, no blur function is applied, but correlated noise is added to the image. The degree of correlation is varied with the application of a symmetric 2D Gaussian kernel with width defined by $\sigma{h_2}$, and noise magnitude $\sigma_n$. These two parameters cover the ranges 0 to 10, and 0 to 0.5, respectively.

\subsubsection{Variable Blur and Noise}
In the last case, both blur and noise degrade the image. Both $h_1$ and $h_2$ are 2D Gaussian kernels with covariance $K$. We parameterize each kernel by considering the eigendecomposition of $K$
$$
K=\left(\begin{matrix}\cos\theta & \sin\theta\\ -\sin\theta & \cos\theta \end{matrix}\right)\left(\begin{matrix}w_1 & 0\\ 0 & w_2 \end{matrix}\right)\left(\begin{matrix}\cos\theta & \sin\theta\\ -\sin\theta & \cos\theta \end{matrix}\right)^{-1}.
$$
We may then define a parameter space for variable blur and noise according to the following derived metrics:
$$ M=\sqrt{w_1 w_2}, \ \ \  E=\ln{\frac{w_1}{w_2}}, \ \ \ 
O=\theta $$
which we label magnitude($M$), eccentricity($E$), orientation($O$). Thus both blur and noise correlation are generally asymmetric with varying extent (magnitude). The range of these parameters is the same for both the blur and noise kernel with specific values given in Table~\ref{tab:scenarios}. overall noise level is again controlled by $\sigma_n$. 

\subsection{Data generation}
In this study, we utilized data from the publicly accessible LIDC dataset\cite{Armato2015Data, Armato2011LIDC, Clark2013TCIA}. We used 40 patient datasets, with 4800 slices from the first 30 patients for training and 1000 slices from the remaining patients for validation. Voxel values were normalized between 0 and 1. To decrease the percentage of zero-value voxel in the CT slices of the training data, we employed a thresholding and hole-filling algorithm to mask the couch and generate a body mask. We ensured all training patch centers were located inside the patients' bodies. In the variable blur and noise scenarios, we created 4800 image realizations with varying blur or noise levels. 
For the combined variable blur and noise scenario, to conserve memory, we produced 4800 realizations for each term in the forward model ($h_1*\tilde{\mu}$ and $h_2*N$) and combined them on the fly. The particular degradation applied to each image is defined by the parameters and ranges specified in Table~\ref{tab:scenarios}.

\subsection{Training Details}
The model was trained end-to-end using the L1 loss function, with a batch size of 64 with the Adam optimizer\cite{adam}, a constant learning rate of $1e^{-4}$, and momentum parameters of $\beta_1=0.9$, $\beta_2=0.999$. For all weight-based and mixed models, $b_w$ was set to 8. In the input-based and mixed model, $b_i$ was set to 8. Each model underwent 5000 epochs of training, as we observed that the loss converged and did not decrease further after this point. The model with best validation loss over 5000 epochs was chosen for testing.

\subsection{Performance Evaluation}
Restoration performance of the proposed approach was evaluated using the peak signal-to-noise ratio (PSNR) between the restored and the ground truth images. For each network architecture, we investigated the model with the best PSNR performance on the validation dataset and evaluated PSNR on the test dataset. The test dataset is made of 10000 image patches sampled from a range of degradation levels. Degradations matched those levels in training that are specified in Table~\ref{tab:scenarios}.

Our performance evaluation includes both a summary PSNR study on the mean performance of each method, but also an investigation of the ensemble performance across many different images and levels of degradation. We additionally form histograms of the relative performance across the test data. Conclusions and representative examples of performance in different performance regions are shown and discussed. Moreover, we consider an "input space" study wherein the performance as a function of degradation parameters is investigated. 

\section{Results}
\begin{table*}[!t]
\centering
\renewcommand\arraystretch{1.5}
\caption{Restoration Performance}
\resizebox{1\textwidth}{!}{%
    \begin{tabular}{m{5cm}<{\centering}|c|c|c|c|c}
    \hline
    \multirow{2}{*}{Scenario}                & \multicolumn{5}{c}{Test PSNR(dB)}           \\ 
    \cline{2-6}
                            & Baseline 1 & Baseline2 & Input-Variant & Weight-Variant & Mixed  \\
    \hline
    Variable Blur           & 30.906     & 30.973    & 30.955        & 31.067         & 31.190 \\
    Variable Noise          & 36.939     & 37.378    & 37.889        & 38.104         & 38.263 \\
    Variable Blur and Noise & 28.843     & 29.488    & 30.383        & 29.625         & 30.685 \\
    \hline
    \end{tabular}%
}
\label{tab:restoration performance}
\end{table*}

\subsection{PSNR Summary}
A summary of restoration performance showing the PSNR between the restored and the ground truth images for each network architecture is shown in Table~\ref{tab:restoration performance}. Note that in all scenarios, the proposed model with auxiliary inputs outperforms its corresponding baseline model. The mixed model outperforms all other models in every scenario. The weight-based model outperforms the input-based model in the variable blur and noise scenario, while it performs worse in the variable blur and noise scenario.

In the variable blur scenario, we observed the smallest improvement in terms of PSNR, possibly because the baseline model can already handle variable blur degradation effectively, as the blur kernel may be easily estimated and utilized for restoration by the model from the measurements. In contrast, the variable blur and noise scenario proved to be the most challenging for the model to estimate the level of degradation from the measurement. Consequently, the most notable improvement for when accurate blur and noise information was provided was in this variable blur and noise scenario. We observed better performance in Baseline 2 compared to Baseline 1, perhaps indicating that part of the improvement is due to the increased number of parameters.

\subsection{Detailed Evaluation}
To further understand the value of including auxiliary information in the neural network performance, we created histograms of PSNR difference between an image restored by the baseline model and our proposed model on the test dataset. We concentrate this analysis on the mixed model since that is where the greatest performance improvement was observed (and since similar observations are present for the input- and weight-variant models). We use Baseline 2 model for comparison, since it has a similar numbers of parameters to the mixed model. Histograms for the three image degradation scenarios are shown in Figure~\ref{fig:histogram}.

As expected from the previous summary PSNR results, we observed that each histogram has a mean larger than zero for all scenarios - meaning, on average, the mixed method outperforms the baseline. However, individual restorations can deviate from this mean performance with the mixed method performing much better than average, and, in some cases, the network with auxiliary parameters can perform worse than the baseline. To further investigate this performance distribution, we studied representative patches selected from the 10th percentile, 90th percentile, and mean histogram bins (representing poor, good, and mean relative performance of the proposed mixed method compared with the baseline method). We consider each degradation scenario individually:

\begin{figure*}[!t]
\centering
\includegraphics[width=6.5in]{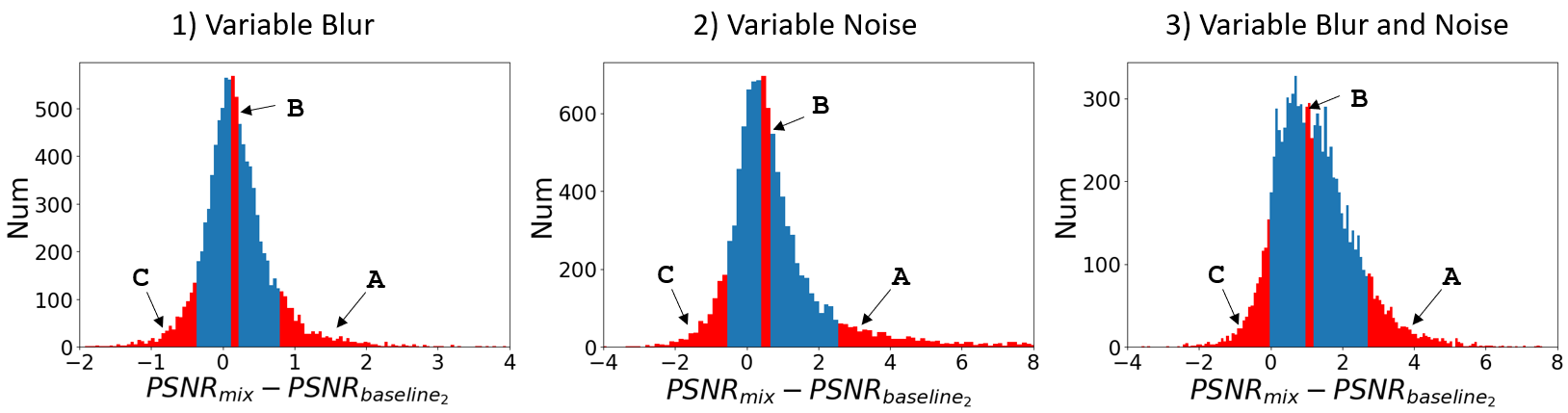} 
\caption{PSNR Difference Histogram between Mixed Model and Baseline 2 across all scenarios. In each chart, regions indicating the top 10\% (A), middle 10\% (B), and bottom 10\% (C) of PSNR improvement are highlighted in red, from which we select the displayed examples.}
\label{fig:histogram}
\end{figure*}

\subsubsection{Variable Blur Scenario}
\begin{figure*}[!t]
\centering
\includegraphics[width=6.5in]{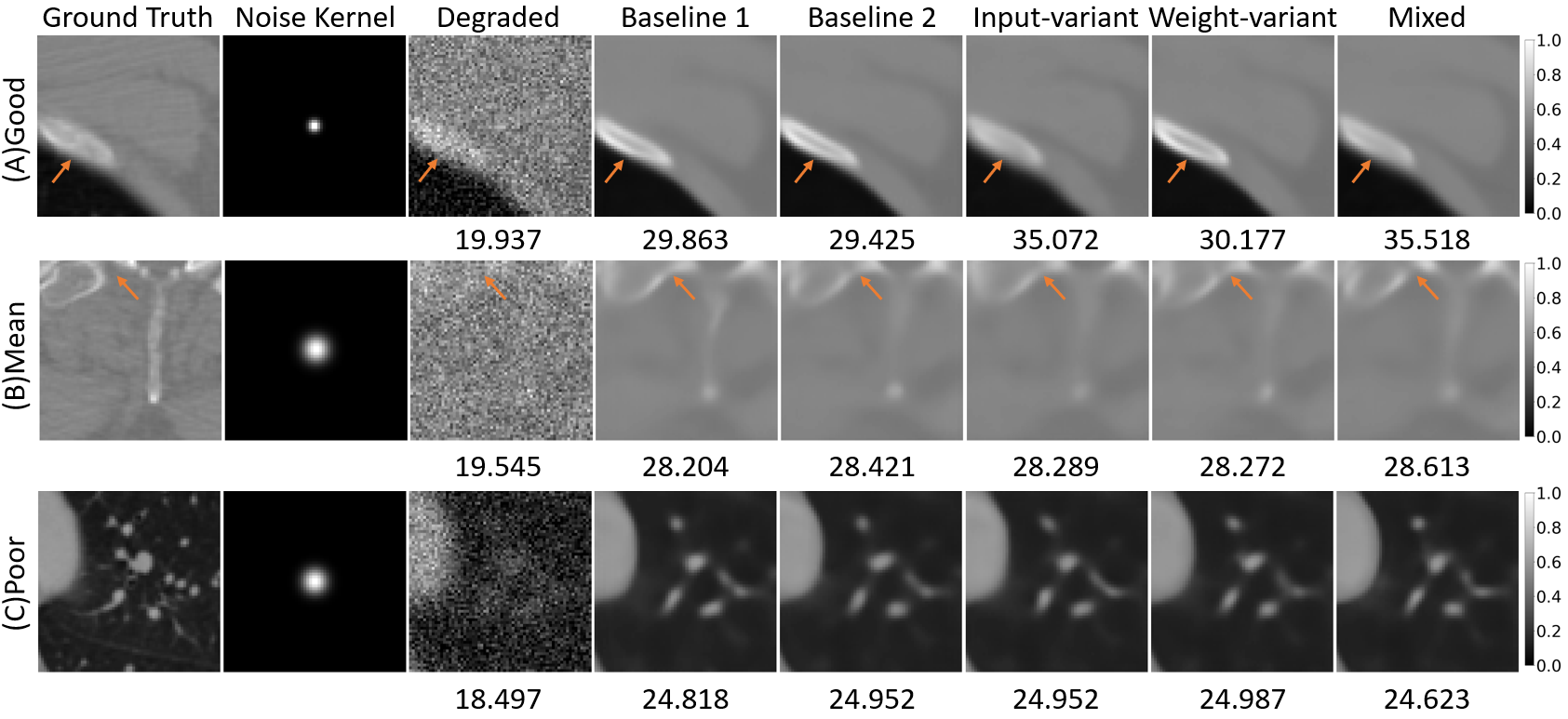}
\caption{Representative restoration examples for the variable blur scenario showing examples of the varied performance across methods. Note systematic oversharpening by baseline approaches in (A); as well as erroneous joining of noncontiguous features in (B).}
\label{fig:examples for variable blur}
\end{figure*}

A representative patch for each performance level in the variable blur scenario is shown in Figure~\ref{fig:examples for variable blur}. We present all network models for comparison. We note a number of observations. Baseline models tended to over-sharpen edges that were not sharp in the ground truth (indicated by the arrow in Figure~\ref{fig:examples for variable blur}A). This suggests that without knowledge of blur the baseline network seeks to sharpen all edges while blur information permits restoration of truly smooth transitions. Similarly, when the blur information was not provided, the baseline model often connected features that were not contiguous (arrow in Figure~\ref{fig:examples for variable blur}B). Again, blur information appears to help with this distinction. A third sample drawn from the poor performance region is shown in Figure~\ref{fig:examples for variable blur}C. We see that important structures were not restored by any of the models. We note that this a relatively high blur case and a patch with many small structures, which presents a significant challenge. The different networks yield different results and varied PSNR; however, no particular trends are observed.

\subsubsection{Variable Noise Scenario}
\begin{figure*}[!t]
\centering
\includegraphics[width=6.5in]{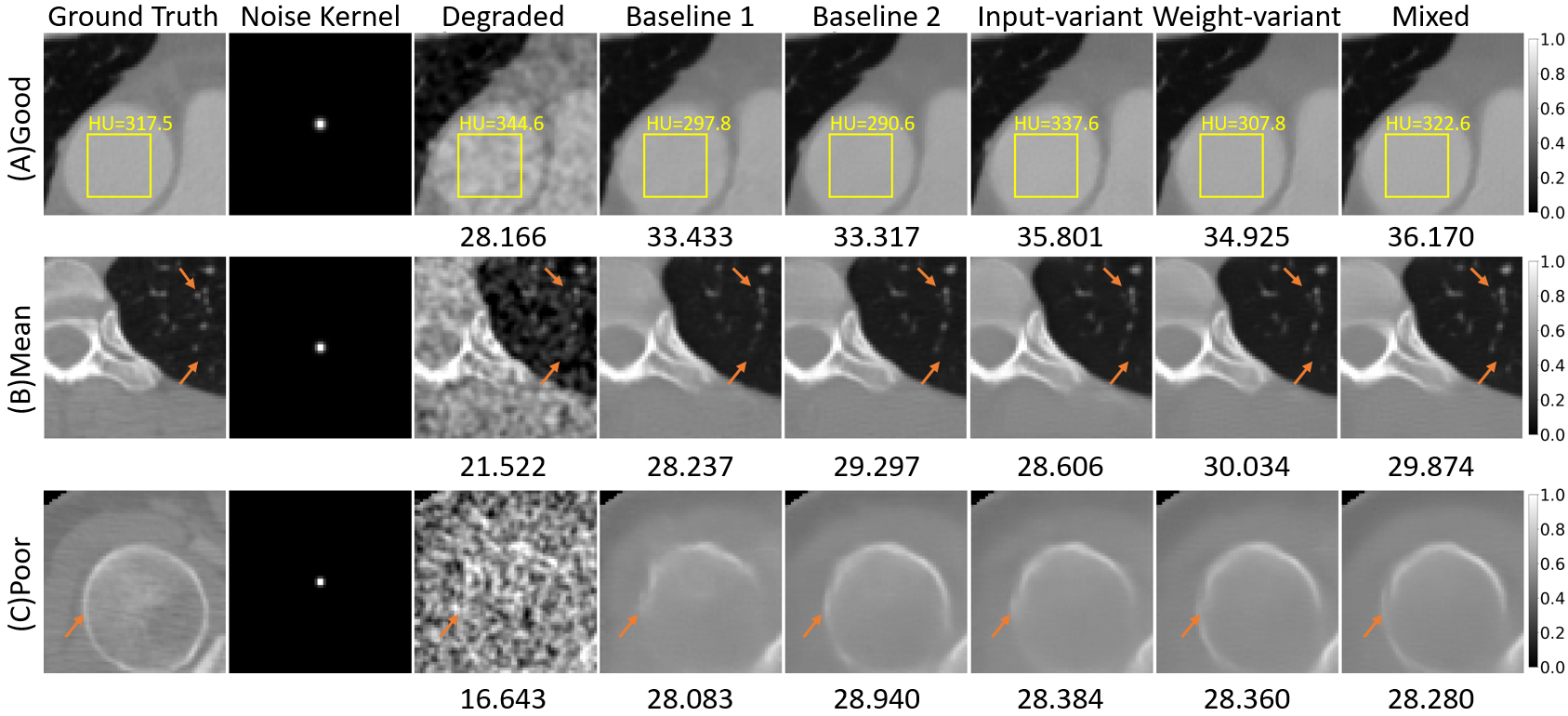}
\caption{Representative restoration examples for the variable noise scenario. In (A) a regional mean attenuation is reported within the labeled square region of interest. Note that the bias in this region is variable across methods with the mixed approach having the lowest error. Small features present in the ground truth appear to be resolved when information about correlated noise is included in the model (B). In very challenging cases (C), it is possible for all models to create plausible but erroneous features.}
\label{fig:examples for variable noise}
\end{figure*}

Representative patches for the variable noise scenario are shown in Figure~\ref{fig:examples for variable noise}. Visually, many restorations are very similar across methods in the "good" performance region. One observed different between approaches is evident in terms of quantitation. To illustrate this, the mean attenuation in a relatively uniform area (indicated by the yellow box in Figure~\ref{fig:examples for variable noise}A) shows different biases across restoration methods. We note that incorporating noise information improves quantitative accuracy. This has a direct impact on PSNR and is potentially important for diagnostic applications involving quantitative biomarkers. In the mean performance sample, we observed that the baseline approaches had difficulty restoring fine detail structures. Conversely, the methods with auxiliary information about the noise distribution appeared to be able to differentiate between structured noise and anatomical features. The example that represents the poor restoration performance is shown in Figure~\ref{fig:examples for variable noise}C. We note that this is a particularly challenging restoration. Relative to Baseline 2, the mixed model does create a larger notch in the bone (non-existent in ground truth) leading to a lower PSNR; however, such errors might be expected in such challenging cases where there are many equally plausible restorations. 

\begin{figure*}[!t]
\centering
\includegraphics[width=6.5in]{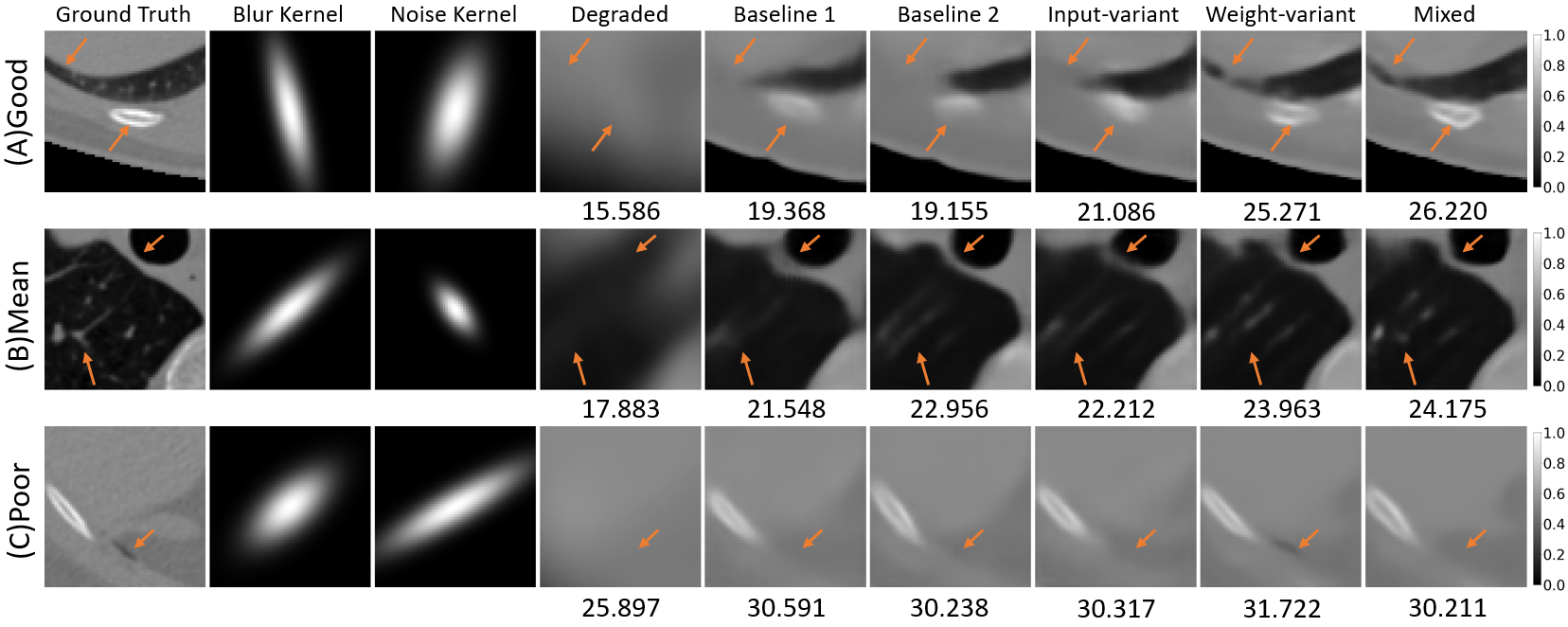}
\caption{Representative restoration examples for the variable blur and noise scenario. The improvements associated with providing auxiliary blur and noise information to the network are particularly evident for anistropic blur functions on images with directional features. While significant improvements are possible, there is still the potential to miss such features in very challenging restoration examples.}
\label{fig:examples for variable blur and noise}
\end{figure*}

\subsubsection{Variable Blur and Noise Scenario}
Sample restorations for the variable blur and noise scenario are shown in Figure~\ref{fig:examples for variable blur and noise}. In the "good" restoration performance example, we see the potential for improvement with auxiliary information. Consider the highly directional blur and noise in Figure~\ref{fig:examples for variable blur and noise}A, vertical spatial frequencies are better restored with the proposed models (see lung region and bone indicated by arrows) when information about those anisotropic properties are provided. Among proposed approaches, the mixed model appears to perform better. Similar observations are seen in the mean performance region (Figure~\ref{fig:examples for variable blur and noise}B), where image features that are obscured in baseline methods can be recovered, even in the directions with the most blur. The methods with auxiliary information are not perfect, however. For challenging examples where the models with auxiliary information cannot resolve some features, though in the case shown in Figure~\ref{fig:examples for variable blur and noise}C, the weight-variant model was able to resolve a fine feature that the mixed model did not. 

\subsubsection{Additional Observations}
\begin{figure*}[!t]
\centering
\includegraphics[width=6.5in]{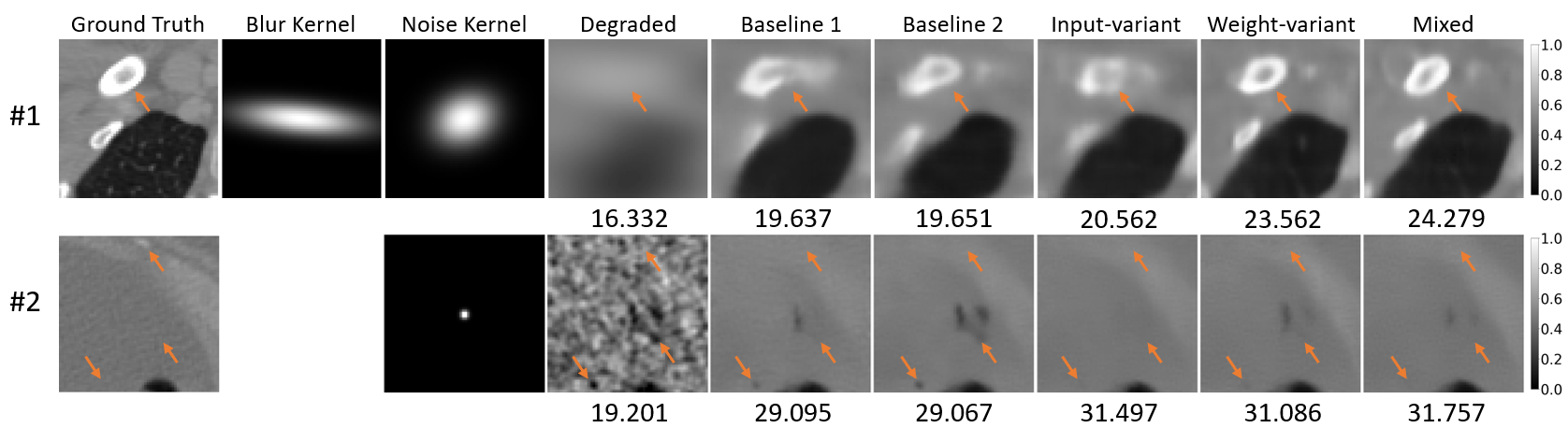}
\caption{Additional examples from the 1: the variable blur and noise scenario, and 2: the variable noise scenario.}
\label{fig:examples for anecdotal observations}
\end{figure*}

In addition to the above examples, we have identified two more cases to highlight the advantages of incorporating blur and noise details into the model. These are shown in Figure~\ref{fig:examples for anecdotal observations}. The first example from the varying blur and noise scenario shows a challenging degradation with a highly anisotropic blur in an image patch with a number of different sized features. The baseline cases have difficulty in restoring the shape of the bony anatomy. As we have seen in other cases (Figure~\ref{fig:examples for variable blur and noise}), as compared with the input-variant model, the weight-variant appears to better handle these challenging directional cases. Though, the mixed model generally outperforms both.

Example 2 draws an example from the variable noise scenario. In this case, the noise kernel imparts correlated noise that appears potentially structural in the degraded image. The baseline models integrate that structure into the restoration. However, those features are better handled in the proposed methods - particularly the input-variant and mixed models. We conjecture that knowledge of the noise correlation prevents such features from coming through in the restoration. Of course, true signals (e.g. the white dot at the top of the patch) that are the same size as the correlation length of the noise, are also not restored. 

\subsection{Input space evaluation}

\begin{figure*}[!t]
\centering
\includegraphics[width=6.5in]{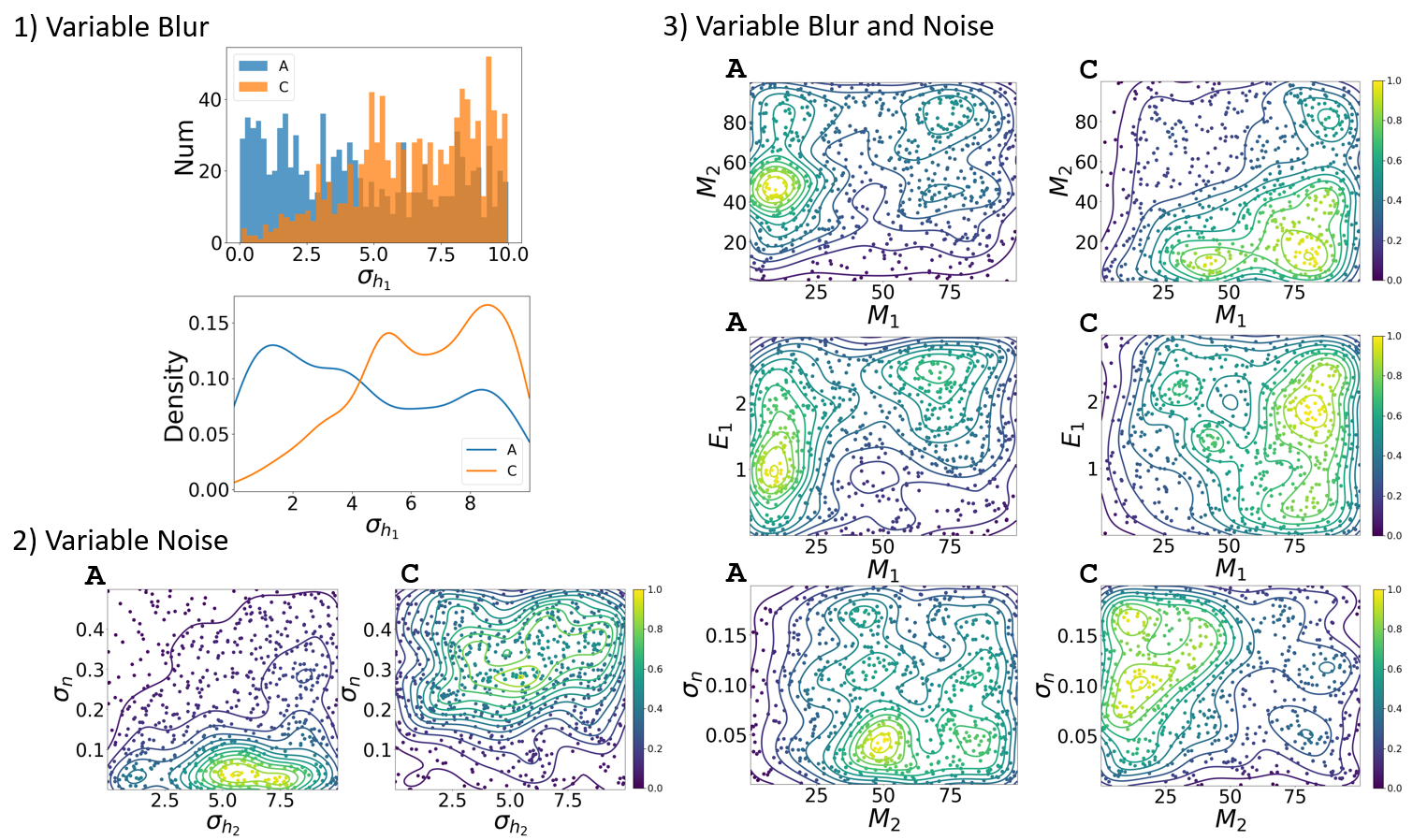}
\caption{Illustration of the performance of the mixed model versus the baseline model (without auxiliary information) as a function of the input parameter space. Specifically, relative performance for each of the three scenarios is shown with density estimates for regions A and C of the Figure~\ref{fig:histogram} histograms. Region A represents situations where the mixed model outperforms the baseline and region C is where the baseline outperforms the mixed model.}
\label{fig:input space evalution}
\end{figure*}

In order to better understand the differences between the baseline model and models that incorporate auxiliary information, we looked at performance as a function of the noise and blur parameters. This "input space" evaluation helped to identify in which cases the auxiliary information yielded the biggest advantage. Again, we focus on the mixed model, as this model performed best and the observations were similar for the other approaches.

For each scenario, we isolated cases from top and bottom 10\% of the PSNR difference histogram. That is, patches and degradations in regions A and C in Figure~\ref{fig:histogram}, representing examples of better performance by the proposed mixed model, and by the baseline model, respectively. We then looked the number of cases in each performance zone as a function of the degradation parameters. For example, for the variable blur scenario, which is governed by a single parameter, ${\sigma_h}_1$, we can make a histogram of the number of cases where the baseline performs best (C) and where the mixed model performs best (A). We also show a smoothed version of the histogram formed by Gaussian kernel density estimation.
For the variable noise and variable blur and noise scenarios, we looked at two-dimensional slices of the degradation parameter space and show contours of the estimated sample densities. A summary of these results for each scenario is shown in Figure~\ref{fig:input space evalution}.

\subsubsection{Variable Blur Scenario}
In the variable blur scenario, we observe that cases where the proposed mixed model performs better are more likely to be in low $\sigma_{h_1}$ regions. Similarly, it is less likely for the baseline approach (without auxiliary information) to outperform the proposed model when $\sigma_{h_1}$ is small. This implies that the auxiliary information is more important for low blur cases. This consistent with previous observations that the baseline method can over-sharpen edges that should be smooth; where the proposed models provide less sharpening based on the knowledge of low blur.

\subsubsection{Variable Noise Scenario}
In the variable noise scenario, we see a significant trend where the proposed model tends to outperform the baseline when the noise level, $\sigma_n$, is lower (as indicated by the higher probably density in  Figure~\ref{fig:input space evalution}-2A). There is also a trend where the mixed model is better for larger noise correlation lengths (e.g. larger $\sigma_{h_2}$). This is a smaller effect than the overall noise level. The cases where the baseline performs much better happen in the opposite situations - higher noise levels and, to a lesser degree, shorter correlation lengths. The observation that the mixed model performs best when noise levels are lower again suggests that the proposed model has an advantage when the auxiliary information helps to moderate the action of the network (e.g. providing less aggressive denoising when the noise levels are low).

\subsubsection{Variable Blur and Noise}
In the third scenario of variable blur and noise, the parameterization of the input space has seven dimensions. A few select 2D slices of this space that show significant trends are illustrated in Figure~\ref{fig:input space evalution}-3.

Looking at $M_1$ versus $M_2$, we observe that the mixed model generally performs best for lower blur, $M_1$, levels at moderate to higher noise levels, $M_2$ (though this effect is somewhat reduced at moderate blur levels). There is a similar, but opposite trend, for the regions where the baseline approach performs better - specifically, at lower noise levels and higher blurs.

For $M_1$ versus $E_1$, the aforementioned trend with respect to $M_1$ is still apparent (mixed model is better for low blur). However, we also observed that good mixed model performance tends to be found in the higher $E_1$ region even when $M_1$ is large. That is, auxiliary information appears to have an impact in cases where the blur is more eccentric. This is consistent with other observations where highly directional blur is handled better by the proposal model. In contrast, we do not observe a strong trend for the cases where the baseline model performs best as a function of eccentricity.

In the $M_2$ vs $\sigma_n$ space, we observed that the mixed model tended to perform best at lower $\sigma_n$ and higher $M_2$. The opposite general trend is apparent. 

\subsubsection{Overall Trends}
In general, the majority of the above observations indicate that the mixed model has a particular advantage over the baseline approach when levels of blur and noise are lower. This suggests that the auxiliary information provided to the network can help to moderate the restoration without over-sharpening or over-denoising. That is not to say the mixed model only has advantage at low noise and blur, since there are enhancements in image quality with the mixed model across noise and blur levels.

\section{Discussion and Conclusion}

In this study, we demonstrate the efficacy of providing a deep learning model with auxiliary inputs to enhance the restoration of images from their degraded versions. The concept is general and may be applied to different network architectures. While one might expect that a network trained specifically for a single blur and noise scenario would outperform our unified model, the proposed approach allows for flexibility in application. That is, the restoration can accommodate changes in noise or blur levels. That the baseline models without auxiliary inputs perform worse on average than the approach with the auxiliary model suggests that it is difficult for the baseline methods to estimate noise and blur directly from the image data. This difficulty appears to increase as the diversity of blur and noise inputs increases. It is possible that a more sophisticated network (deeper architecture and more training) would have improved performance, and that the performance advantage with auxiliary inputs lies in efficiency. Such studies are future work.


In this work, relatively simple noise and blur models were used. Blur in CT images is often complex, and can exhibit anisotropic and shift-variant characteristics. Sources of blur include motion blur caused by gantry rotation during detector integration, and x-ray source blur influenced by the size of the focal spot. Patient-dependent blur can arise from anatomical motion as well as statistical reconstruction methods where blur is influenced by noise level. CT image noise is generally non-stationary and image-dependent (e.g. larger patients, bony anatomy, etc. attenuate more x-rays, leading to higher local noise and significant noise correlations. In this study, we partially account for the anisotropic property of blur and the correlation of noise by adopting bi-variate Gaussian kernels as rough approximations. Future work will consider noise and blur models with additional degrees of freedom that better represent the diversity of image properties seen in CT. Similarly, the restoration process presented here is applied in a locally shift-invariant fashion. Such processing is appropriate for many reconstruction methods that are linear and/or locally linearizable (e.g., FDK reconstruction for blur, penalized-likelihood reconstruction with a quadratic penalty for blur and noise, etc.). Incorporating such realism, extension to shift-variant methods, and translation to physical data is the subject of ongoing work.

\section{Disclosures}
The authors have no relevant financial interests or conflicts of interest to disclose.

\section{Code, Data, and Materials Availability}
Research code for this work is not currently available. Data used in this work was from the publicly accessible LIDC dataset.\cite{Armato2015Data, Armato2011LIDC, Clark2013TCIA}.

\vspace{-0.1in}
\section{Acknowledgments}
This work was supported, in part, by NIH grant R01EB031592.


\bibliography{report}   
\bibliographystyle{spiejour}   

\end{spacing}
\end{document}